\documentclass[aps,prl,showpacs,twocolumn]{revtex4}
\usepackage{graphicx}
\usepackage{amssymb}
\usepackage{amsmath}
\usepackage{bm}
\pdfoutput=1
\begin{document}

\title[]{Probing Phase Fluctuations in a 2D Degenerate Bose Gas by Free Expansion}

\author{Jae-yoon Choi}
\author{Sang Won Seo}
\author{Woo Jin Kwon}
\author{Yong-il Shin}\email{yishin@snu.ac.kr}

\affiliation{Center for Subwavelength Optics and Department of Physics and Astronomy, Seoul National University, Seoul 151-747, Korea}

\date{\today}

\begin{abstract}
  We measure the power spectrum of the density distribution of a freely expanding 2D degenerate Bose gas, where irregular density modulations gradually develop due to the initial phase fluctuations in the sample. The spectrum has an oscillatory shape, where the peak positions are found to be independent of temperature and show scaling behavior in the course of expansion. The relative intensity of phase fluctuations is estimated from the normalized spectral peak strength and observed to decrease at lower temperatures, confirming the thermal nature of the phase fluctuations. We investigate the relaxation dynamics of nonequilibrium states using the power spectrum. Free vortices are observed with ring-shaped density ripples in a perturbed sample after a long relaxation time.
\end{abstract}


\maketitle

 Phase coherence is one of the main characteristics of superfluidity and critically affected by the dimensions of the system. In low dimensional systems, large thermal and quantum phase fluctuations prohibit the establishment of long-range phase coherence which is a typical order parameter for a three-dimensional superfluid~\cite{MerminHo_PRL,Hohen_PRL}. Nevertheless, a two-dimensional (2D) interacting system can undergo a superfluid phase transition at a finite critical temperature with an algebraically decaying coherence. This transition is successfully described in the Berezinskii-Kosterlitz-Thouless (BKT) theory~\cite{B_JETP,KT} as a topological phase transition, where the critical point is associated with spontaneous pairing of free vortices with opposite circulations.

 Recent experiments with 2D atomic Bose gases have demonstrated that this BKT physics can be studied in a finite-size trapped sample~\cite{ZH_nature,Kruger07,preSF_PRL,Scale_nature,preSF_PRA,EqnRb_PRL,tri_PRL}. Phase coherence and thermodynamic properties have been investigated using matter-wave interference and by detailed analysis of the \textit{in situ} density and momentum distributions of trapped samples, observing the algebraic decay of coherence~\cite{ZH_nature}, a presuperfluid regime~\cite{tri_PRL,preSF_PRL,Scale_nature,preSF_PRA}, and the scale invariance of the equation of state~\cite{Scale_nature,EqnRb_PRL}. It is now highly desirable to have quantitative probes directly sensitive to phase fluctuations in order to study the topological nature of the phase transition. In particular, nonequilibrium phase dynamics near the critical point would provide valuable insights on the BKT transition. Relaxation dynamics in the 2D XY model have been under intense theoretical investigation~\cite{Yurke93,Relax_PRL} and recently an experimental scheme to study a dynamic BKT transition in 2D Bose gases was proposed~\cite{DynamicalKT}.

 In this paper, we demonstrate a new quantitative probe for phase fluctuations in a 2D degenerate Bose gas using the density correlations in a freely expanding sample. Irregular density modulations gradually develop in the coherent part of the sample during expansion. We observe that the power spectrum of the density distribution has an oscillatory shape and find that the peak positions are independent of temperature and show scaling behavior in the course of expansion. The relative intensity of phase fluctuations is estimated from the spectral peak strength normalized with the central density of the coherent part and we show that it decreases at lower temperature in thermal equilibrium. This confirms the thermal nature of phase fluctuations in the 2D system. In addition, we investigate the relaxation dynamics of nonequilibrium states by measuring the time evolution of the relative intensity of phase fluctuations. Interestingly, ring-shaped density ripples stochastically appear in a perturbed sample after a long relaxation time, which we identify with free vortices having a long lifetime in a 2D sample.

 We prepare a 2D degenerate Bose gas of $^{23}$Na atoms in a single pancake-shape optical dipole trap~\cite{QGLSkyrmion,Choi12}. Thermal atoms in the $|F=1,m_F=-1\rangle$ state are loaded from a plugged magnetic trap~\cite{SNUBEC} into the optical trap and evaporative cooling is applied by reducing the trap depth. The sample temperature is controlled by the final trap depth in the evaporation, resulting in 0.6 to $1.3\times 10^6$ atoms in a sample. Finally, the optical trap depth ramps up and the trapping frequencies $(\omega_x, \omega_y, \omega_z) = 2\pi \times$(3.0, 3.9, 370)~Hz. The cooling procedure is intentionally set to be slow over 15~s, ensuring thermal equilibrium. The lifetime in the optical trap is over 50~s. In the Thomas-Fermi approximation, the chemical potential is about $h\times260$~Hz less than the confining energy $\hbar \omega_z$, so we expect 2D physics in the phase coherence of the sample at low temperature. The dimensionless interaction strength $\tilde{g}=a \sqrt{8\pi m \omega_z/\hbar}\simeq 0.013$, where $a$ is the 3D scattering length and $m$ is the atomic mass. The in-plane density distribution $n(x,y)$ is measured by taking an absorption image after an expansion time $t_e$ which is initiated by suddenly turning off the trapping potential.

\begin{figure}
\includegraphics[width=7.5cm]{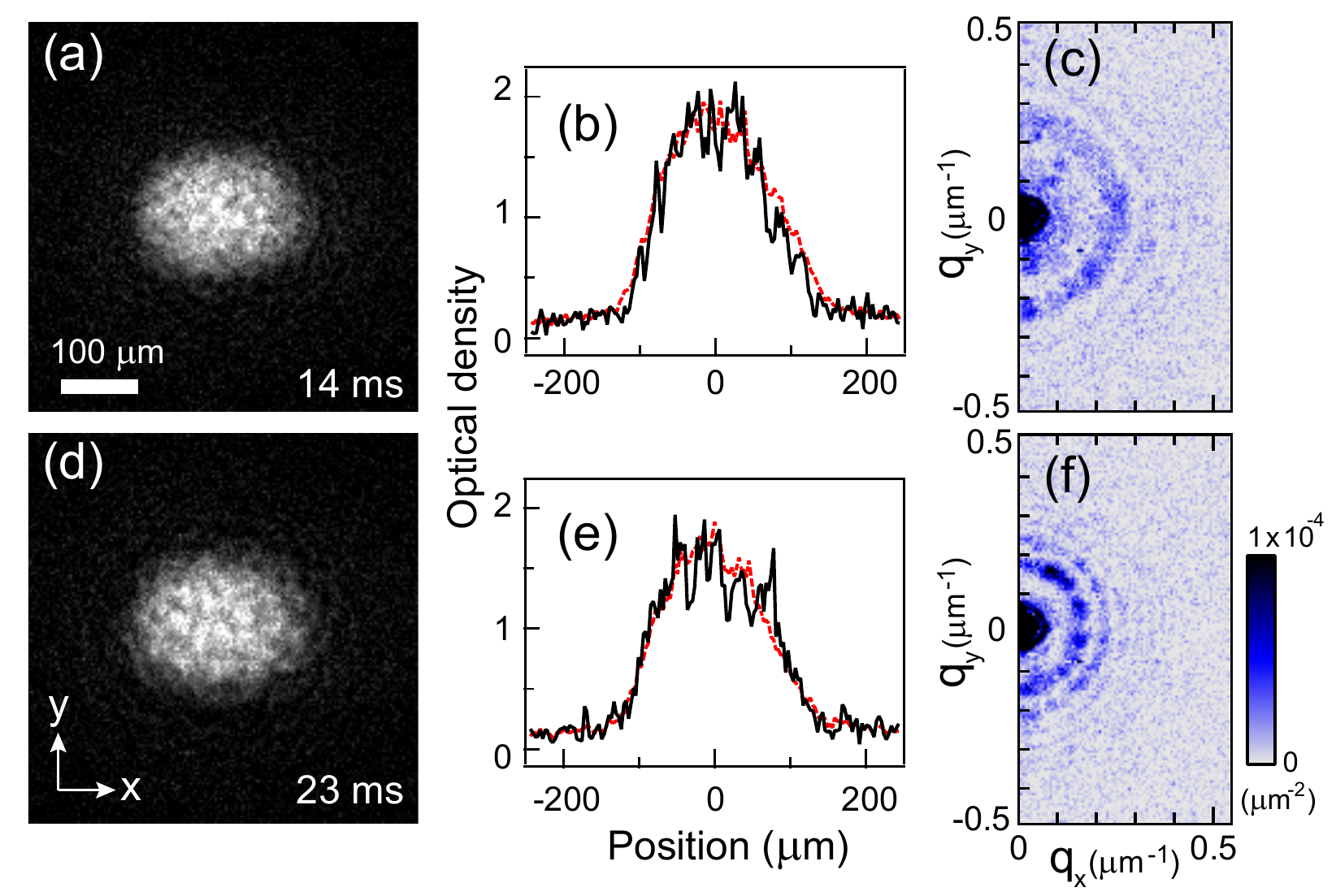}
\caption{
   (color online) Emergence of density fluctuations in a freely expanding 2D Bose gas. Density distributions after (a) $t_e=14$~ms and (d) 23~ms of time-of-flight. Density fluctuations gradually develop during expansion, increasing their length scale and visibility. (b,e) The horizontal density profiles in the center of the samples. The red dashed lines indicate the averaged profiles over 10 individual realizations of the same experiment. The coherent fraction $\eta\approx30\%$ (see text for details). The power spectrum of the density distribution is measured with the magnitude square of its Fourier transform. (c,f) The averaged power spectra corresponding to (a,d).
   }
\label{overview}
\end{figure}

 Expansion has been a conventional and powerful method in quantum gas experiments to study coherence properties of a sample~\cite{tri_PRL,preSF_PRA}. In our experiment, we are interested in the short expansion regime where $t_e\ll 1/\omega_{x,y}$ and the phase coherence information would be revealed as density correlations. Since the sample initially expands fast along the tight direction, the atom interaction effects are rapidly reduced and the subsequent evolution in other directions can be described as free expansion. When the sample contains phase fluctuations, self interference would result in density modulations in the expanding sample. This method has been exploited in previous studies of phase fluctuations in elongated Bose-Einstein condensates~\cite{OneDfluc_PRL} and 1D Bose gases~\cite{oneD_g1_PRA}.

\begin{figure}
\includegraphics[width=8.5cm]{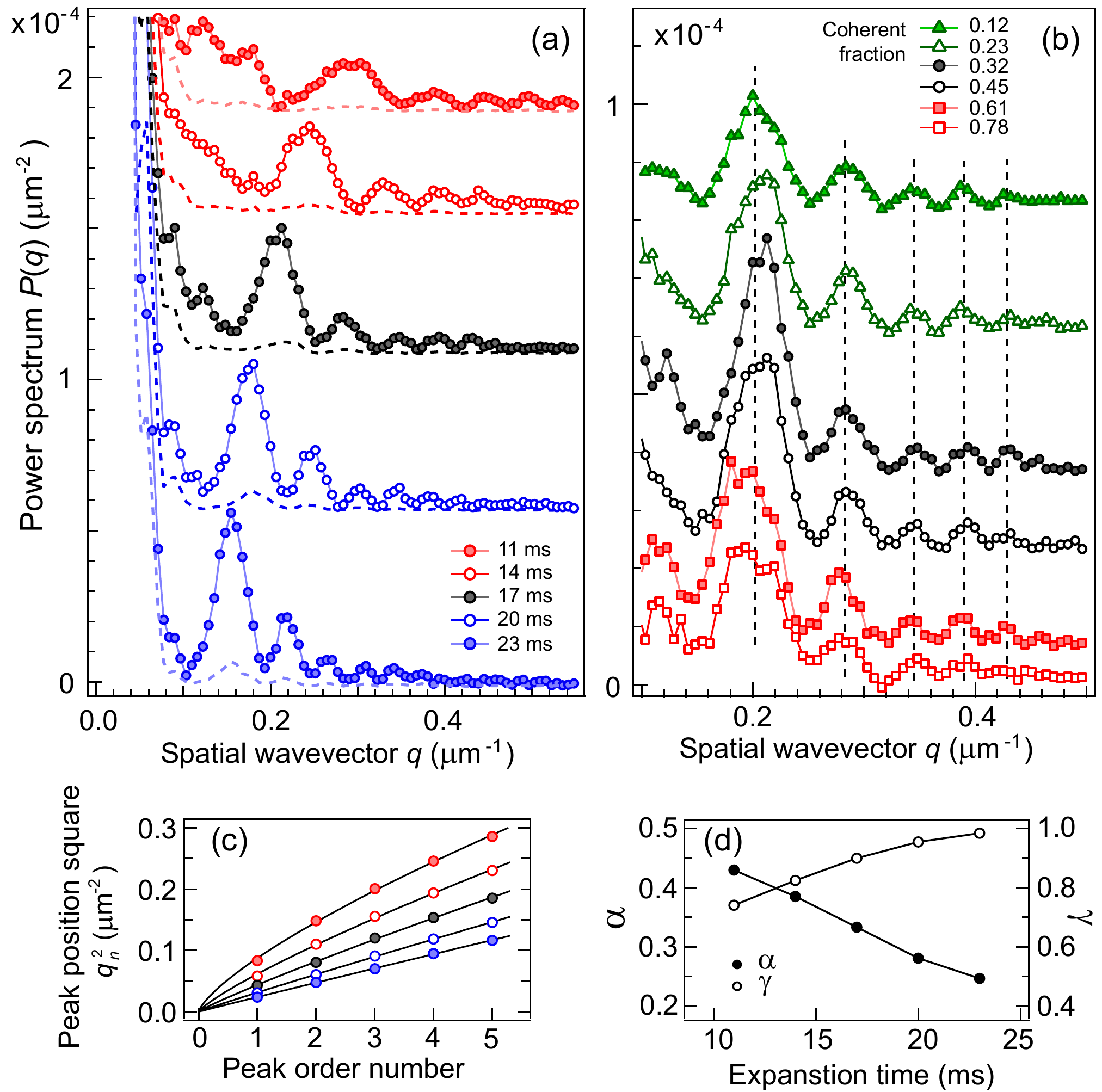}
\caption{
   (color online) The power spectrum of density fluctuations. (a) 1D power spectra $P(q)$ are obtained by azimuthally averaging the averaged 2D spectra for various expansion times $t_e$. Each spectrum is displayed with an offset for clarity. The dashed lines are the corresponding spectra for the averaged density distributions over 10 individual realizations. (b) $P(q)$ at $t_e=17$~ms for various temperatures. The spectral peak positions $q_n$'s are determined from a fit of multiple gaussian curves to each spectrum ($n$ is the peak order number). The vertical dashed lines indicate the average peak positions. (c) The empirical function $\hbar q_n^2 t_e /2\pi m = \alpha n^\gamma$ are fit to the peak positions in (a) for each $t_e$. (d) The expansion time dependence of the two free parameters $\alpha$ and $\gamma$.
   }
\label{Radialavg}
\end{figure}

 We observe that density fluctuations develop in an expanding 2D Bose gas (Fig.~\ref{overview}), where the characteristic size and the visibility of the density lumps increases with the expansion time. Density fluctuations appear discernible only when the sample shows a bimodal density distribution so we refer the center part as the coherent part of the sample in the following~\cite{tri_PRL,Kruger07}. In order to obtain the density correlation information, we measure the power spectrum of the density distribution as the square of the magnitude of its Fourier transform, $P(\vec{q})=|\int dx dy e^{i\vec{q}\cdot\vec{r}}n(\vec{r})|^2$. Although the spatial pattern of the density modulations appears random in each realization, the power spectrum clearly reveals a multiple ring structure that scales down with the expansion time [Fig.~\ref{overview}(c) and (f)]. For quantitative analysis, we obtain an 1D spectrum $P(q)$ by azimuthally averaging the 2D spectrum (Fig.~\ref{Radialavg})~\cite{footnote3}.  The strong signal around $q=0$ corresponds to the finite size of the coherent part.

 The scaling behavior of the spectrum can be qualitatively understood in terms of the Talbot effect~\cite{Talbot,fTalbot}. It is well known in near-field diffraction that when a grating is illuminated by monochromatic waves, the identical self-image of the grating is formed at a distance $L_\textrm{T}=2d^2/\lambda$ away from the grating, where $d$ is the grating period and $\lambda$ is the wavelength of the incident radiation. The same effect occurs with a phase grating~\cite{Lohmann90}. In matter wave optics, $\lambda=h/m\upsilon$, where $\upsilon$ is the incident speed of atoms~\cite{Chapman95}, so the propagation time for self-imaging is defined as $t_\textrm{T}=L_\textrm{T}/\upsilon=2 m d^2 / h$ independent of $\upsilon$. If we consider a 2D Bose gas as a macroscopic matter wave containing phase fluctuations at all length scales, it is expected that the component of wavenumber $q$ satisfying the Talbot condition $q^2= 4\pi m/\hbar t_e$ will emerge predominantly in the density distribution at a given expansion time $t_e$. The multiple peaks in $P(q)$ can be accounted for by the fractional Talbot effect where self-images with smaller periods $d/n$ ($n>0$ is an integer) are produced at $L_\textrm{T}/2n$~\cite{fTalbot}.

 Recently, theoretical calculations on the spectrum of density modulations have been performed for a homogeneous 2D Bose gas at low temperatures~\cite{DensityRipple}, showing that the $n$th peak position $q_n$ closely satisfies $\hbar q_n^2 t_e/ 2\pi m \approx (n-1/2)$. In particular, they predict that for sufficiently long expansion times the spectrum remains self-similar during expansion and its shape is determined only by the exponent of the power-law decay of the first order coherence function.

 In our experiment, we observe that the spectrum preserves its oscillatory shape during expansion and that the peak positions $q_n$'s are independent of temperature (Fig.~\ref{Radialavg}), which are in qualitative agreement with the theoretical prediction. However, we find different scaling behavior of $q_n$'s in the measured spectra, which is well described as $\hbar q_n^2 t_e /2\pi m = \alpha n^\gamma$ with $0.2<\alpha<0.45$ and $0.7<\gamma<1$ for $t_e=10\sim25$~ms [Fig.~\ref{Radialavg}(d)]. Furthermore, the phase of the spectral oscillation is opposite to the theoretical prediction, suggesting that an additional peak is present at $q_0=0$, which is also hinted by the shoulder-like hump in $q<q_1/\sqrt{2}$. We rule out the finite size effects by seeing no dependence of $q_n$'s on sample size as well as temperature, implying that the observed scaling behavior might be intrinsic to the expansion dynamics. We note that the long expansion time condition $\sqrt{\hbar t_e/m} \gg \xi_{2D}$ for the validity of the theoretical prediction is marginally fulfilled in our experiment ($\sqrt{\hbar t_e/m}/\xi_{2D}\sim )$, where $\xi_{2D}=\hbar/\sqrt{m\mu}$ is the 2D healing length, $\mu$ being the chemical potential~\cite{footnote4}.

\begin{figure}
\includegraphics[width=7.5cm]{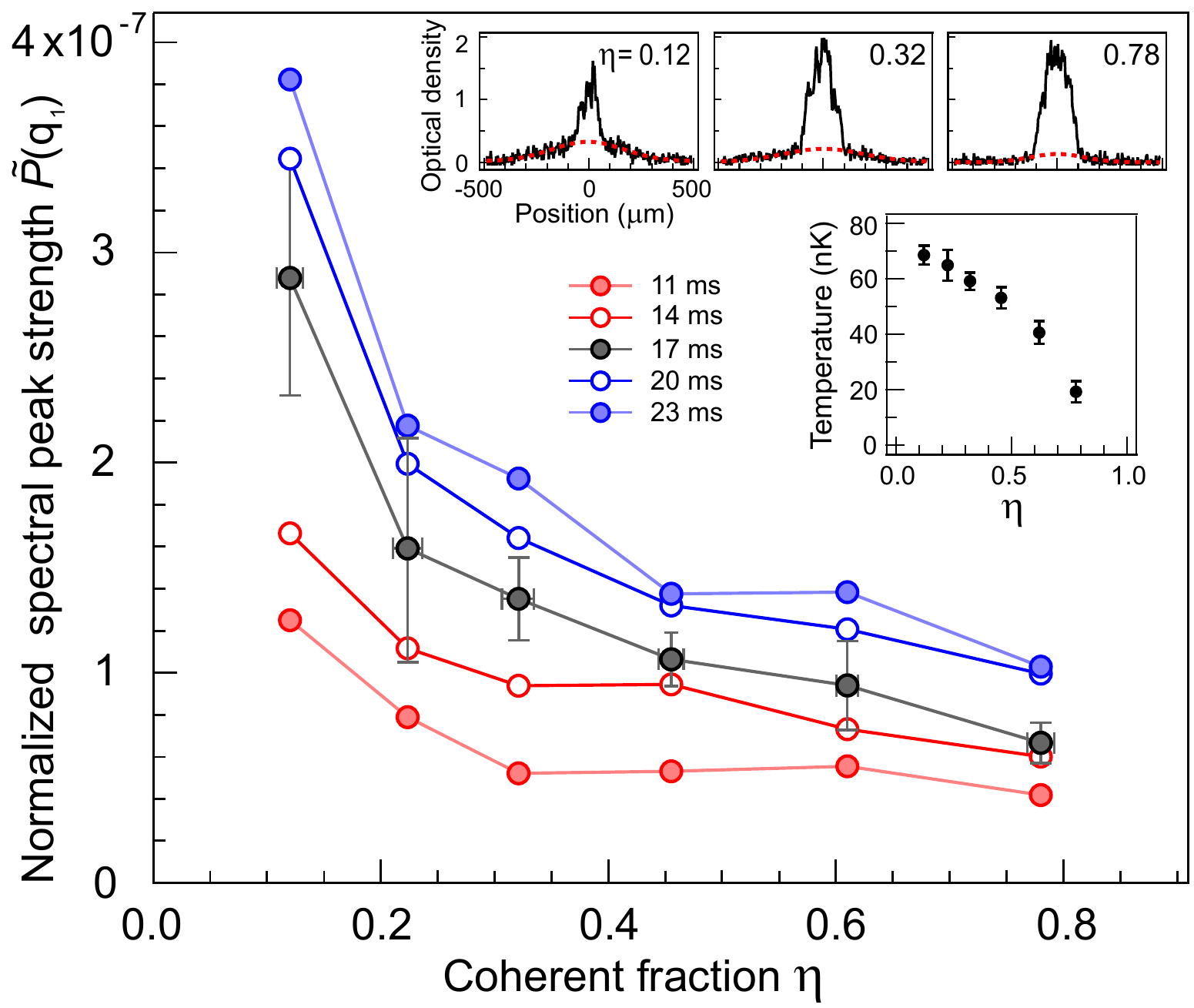}
\caption{
   (color online) Temperature dependence of phase fluctuations at thermal equilibrium. The relative intensity of phase fluctuations is estimated from the normalized strength of the first spectral peak $\tilde{P}(q_1)$ and the relative temperature is parameterized with the coherent fraction. The upper insets show the density profiles for different temperatures. The red dashed lines are gaussian fits to the thermal wings, from which temperatures are estimated (lower inset). Each data point consists of ten independent measurements and error bars indicate standard deviation.}
\label{PeakHvsT}
\end{figure}

 The universality of the spectral peak positions suggests that the spectral peak strength can be used as a measure of the magnitude of phase fluctuations in a sample. In order to quantify the relative intensity of phase fluctuations, we normalize the strength of the first spectral peak with the square of the central density $n_c$ of the coherent part in the sample, $\tilde{P}(q_1)=P(q_1)/n_c^2$, where $n_c$ is determined from a fit of two gaussian curves to the density distribution~\cite{Kruger07,tri_PRL,Hadzibabic08,Holzmann08,footnote1}.

 Using $\tilde{P}(q_1)$, we first investigate the temperature dependence of phase fluctuations in thermal equilibrium. To estimate the relative temperature to the critical point in a model-independent way~\cite{preSF_PRL,EqnRb_PRL}, we use the coherent fraction $\eta$ that is defined as the ratio of the atom number of the coherent part to the total atom number. The value of $\eta$ was constant within 5$\%$ for our expansion times. Fig.~\ref{PeakHvsT} shows that $\tilde{P}(q_1)$ is suppressed at lower temperature (higher $\eta$), confirming the thermal nature of phase fluctuations.

 For a weakly interacting 2D Bose gas, especially for our small $\tilde{g}=0.013$, the BKT critical temperature $T_c$ is close to the BEC critical temperature for a trapped ideal Bose gas $T_{c,\textrm{BEC}}=0.94 \hbar(\omega_x \omega_y \omega_z N)^{1/3}/k_B$~\cite{tri_PRL,Hadzibabic08,Holzmann08}. We estimate $T_c \approx 80$~nK for $N\approx 1.3\times 10^6$. Since $k_B T_c \approx 4\hbar \omega_z$, thermal populations in the tight direction is not negligible, accounting for the gaussian-like profile of the saturated thermal cloud~\cite{Hadzibabic08,Holzmann08}. The rapid increase of $\tilde{P}(q_1)$ at $\eta< 0.2$ might indicate the behavior in the proximity of the critical point. In Ref.~\cite{tri_PRL}, the critical point was identified at $\eta\approx0.1$ with the abrupt change in the width of the coherent part. Since the spatial extent of the coherent part becomes small, the signal-to-noise ratio is poor when $\eta \leq 0.05$ so we cannot study the presuperfluid regime where the decay of the coherence function changes from algebraic to exponential, which might be reflected in the spectral shape.

\begin{figure}
\includegraphics[width=8.5cm]{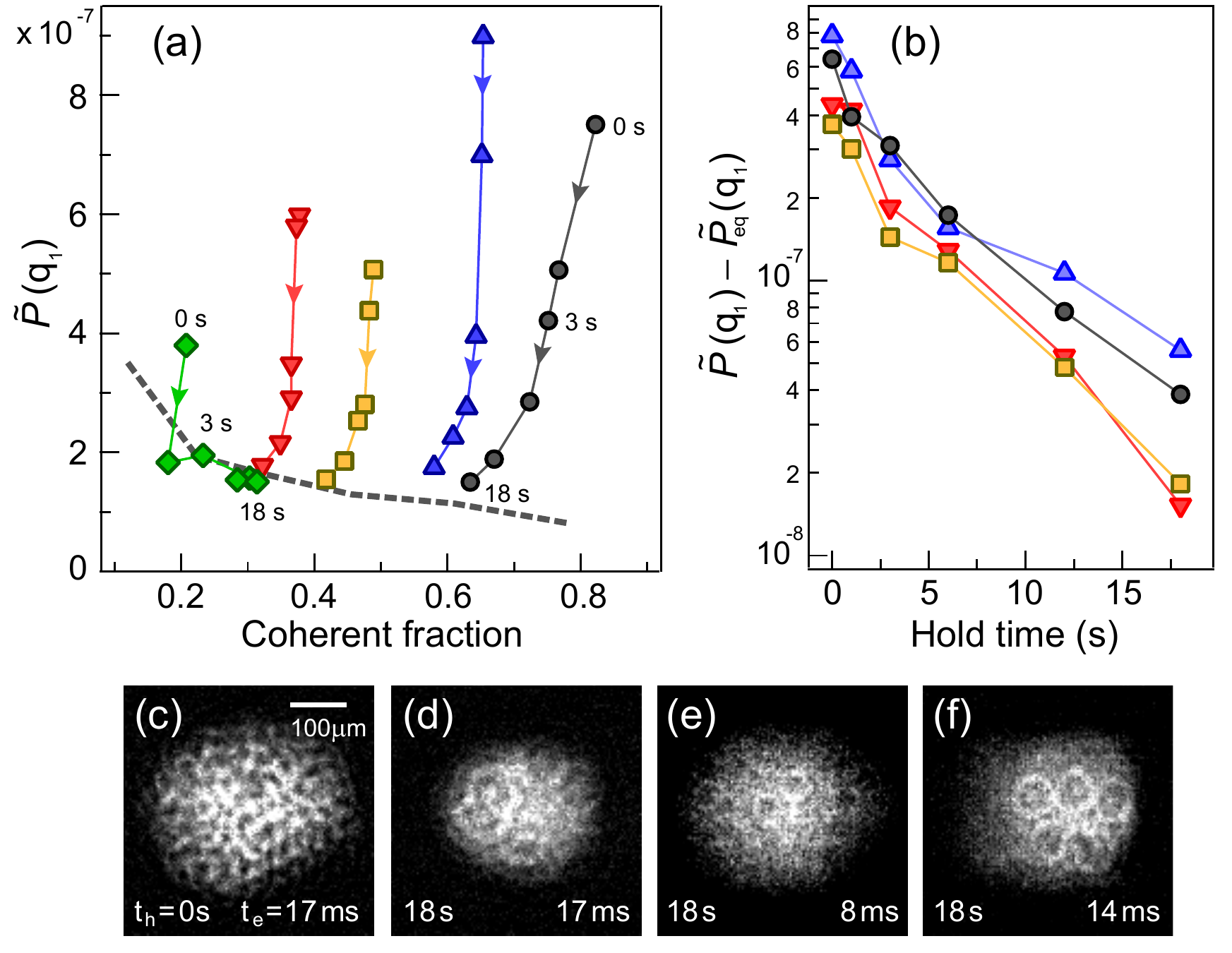}
\caption{
   (color online) Relaxation of a nonequilibrium 2D Bose gas. (a) Time evolutions of various nonequilibrium states in the plane of $\tilde{P}(q_1)$ and $\eta$ ($t_e=17$~ms). The dashed line is the interpolation for the thermal equilibrium data with $t_e=17$~ms in Fig.~\ref{PeakHvsT}. (b) The decay curves of the excess phase fluctuations $\tilde{P}(q_1)-\tilde{P}_\textrm{eq}(q_1)$, where $\tilde{P}_\textrm{eq}(q_1)$ is the equilibrium value corresponding to $\eta$ at $t_h=18$~s of hold time. Density distributions of the lowest temperature sample at (c) $t_h=0$~s and (d) 18~s with $t_e=17$~ms, and at $t_h=18$~s with (e) $t_e=8$~ms and (f) 23~ms. The samples in (e) and (f) were strongly perturbed to see more ring-shaped density ripples.
   }
\label{Relaxation}
\end{figure}

 The power spectrum can be used to study nonequilibrium dynamics in a 2D Bose gas. For this study, we prepare a 2D sample in a nonequilibrium state by transferring a condensate instead of thermal atoms from the plugged magnetic trap into the optical trap. The induced perturbations are small enough that the density profile in the optical trap is quite close to that at equilibrium. In Fig.~\ref{Relaxation}(a), we plot the time evolutions of samples in various initial conditions in the plane of $\tilde{P}(q_1)$ and $\eta$ (Fig.~\ref{Relaxation}f), clearly showing that the nonequilibrium states decay to equilibrium. The decay time of the excess phase fluctuations with respect to the equilibrium value is measured to $\sim 4$~s, corresponding to $\sim 10$ collision times in our typical condition. Note that the hottest sample first decays and then moves along the equilibrium line with increasing $\eta$ because of the evaporation cooling due to the finite trap depth. This verifies that our previous measurements are indeed for phase fluctuations in thermal equilibrium.

 Remarkably, ring-shaped density ripples are observed in the perturbed samples after long relaxation times (Fig.~\ref{Relaxation}). We believe that this corresponds to vortex excitations generated in the sample transferring procedure. Since a vortex in 2D can decay only via pairing with another vortex with opposite circulation or drifting out of the finite sample, we may expect a metastable state with vortices having a long lifetime. Recently, it has been reported that vortex excitations survive longer in an oblate condensate because Kelvin mode excitations on a vortex line are suppressed~\cite{VortexDiplole_PRL,KelvinSuppress_PRA}. The ring pattern appears more often at lower temperature with stronger perturbations. In the two lowest temperature cases (circle and triangle in Fig.~\ref{Relaxation}), the appearance probability is about 60\% at $t_h=18$~s of hold time, where the spectral strength of the samples with vortices is about 10\% higher than without them. In Fig.~\ref{Relaxation}(b), the decay rates of the two coldest samples become slightly slower after $t_h>5$~s, which might be attributed to the long lifetime of vortices. Fig.~\ref{Relaxation}(d)-(f) shows the expansion dynamics of the ring-shape density ripples. In the thermal equilibrium case, the ring pattern was never seen for $\eta>0.2$.

 In conclusion, we have demonstrated the power spectrum of the density fluctuations in a freely expanding 2D Bose gas as a new quantitative probe for phase fluctuations. Together with more controlled perturbations~\cite{Ferrodomain,Decohere_Nature}, we expect this method to be extended for studying nonequilibrium phenomena in BKT physics such as critical exponents~\cite{Yurke93,Relax_PRL} and dynamic transitions~\cite{DynamicalKT,Roumpos12}.

 This work was supported by the NRF grants funded by the Korea government (MEST) (Nos. 2010-0010172, 2011-0017527, 2008-0062257, and WCU-R32-10045). JC, SSW, and YS acknowledge support from the Global PhD Fellowship, the Kwanjeong Scholarship, and the TJ Park Science Fellowship, respectively.


\begin{references}
\bibitem{footnote3} The resolution of our imaging system is about 5~$\mu$m, so the spectral signal at $q>1~\mu$m$^{-1}$ is purely contributed from photon shot noises. We subtract the constant offset value at high $q$ from the spectrum, which is typically a few $\%$ of $P(q_1)$.
\bibitem{footnote4} In our experiment, $\sqrt{\frac{\hbar t_e}{m}}/\xi_{2D}=\sqrt{t_e \mu / \hbar} <6$. The focal depth of our imaging system is $\sim 100~\mu$m, limiting the expansion time $t_e<40$~ms.
\bibitem{footnote1} At low temperatures, the density profile of the coherent part is close to a parabola shape, but the gaussian fit still gives a good estimation for $n_c$ within a few $\%$.
\end{references}
\end{document}